\documentclass[useAMS,usenatbib,referee]{mn2e}
\usepackage{epsfig}

\title{Torsional shear oscillations in the neutron star crust driven by
restoring force of elastic stresses}
\author[S. I. Bastrukov,
  H.-K. Chang, J. Takata, G.-T. Chen
  and I. V. Molodtsova]{S. I. Bastrukov,$^{1,2}$
  H.-K. Chang,$^1$ J. Takata,$^3$ G.-T. Chen$^1$
  and I. V. Molodtsova$^2$\\
  $^1$ Department of Physics and  Institute of Astronomy,
  National Tsing Hua University, Hsinchu, 30013, Taiwan\\
 $^2$ Laboratory of Informational Technologies, Joint Institute for
 Nuclear Research,
141980 Dubna, Russia\\
 $^3$ ASIAA/National Tsing Hua University - TIARA, Hsinchu, 30013,
 Taiwan}
\begin{document}

 \maketitle

\label{firstpage}

\begin{abstract}
 We present several exact solutions of the eigenfrequency problem for
 torsional shear vibrations in homogeneous and non-homogeneous
 models of the neutron star crust governed by canonical equation of
 solid mechanics with a restoring force of Hookean elasticity.
 Particular attention is given to regime of large lengthscale
 nodeless axisymmetric differentially rotational oscillations which
 are treated in spherical polar coordinates reflecting real geometry of the neutron star crust.
  Highlighted is the distinction between analytic
   forms and numerical estimates of the frequency, computed
   as a function of multipole degree of nodeless torsional oscillations and fractional depth
   of the crust, caused by different boundary conditions imposed on the toroidal field of material displacements.
   The relevance of considered models to quasiperiodic oscillations, recently detected during the flare of SGR 1806-20
   and SGR 1900+14, is discussed.
\end{abstract}


\section{Introduction}
 Ever since the identification of pulsars with neutron stars, the non-radial torsional shear oscillations
 restored by bulk forces of different in physical nature internal stresses have been and still are among the
 most important issues in the study of the interconnection between the electromagnetic activity and
 asteroseismology of pulsars
  (e.g. Ruderman 1969, van Horn 1980; Hansen and Cioffi 1980;
  Schumaker and Torne 1983; McDermott,  van Horn \& Hansen 1988, Strohmayer 1991; Duncan 1998; Bastrukov,  Weber,
  \& Podgainy, 1999;
  Yoshida, Lee 2002,
  Bastrukov, Podgainy, Yang \& Weber, 2002;  Bastrukov, Chang, Mi\c sicu \c S, Molodtsova \& Podgainyi 2007). Recently,
  increasing interest in this domain of research has been prompted by the discovery of quasiperiodic oscillations
  (QPOs) on the lightcurve tail of SGR 1806-20 (Izrael et al 2005) and SGR 1900+14 (Strohmayer \& Watts
  2006). It has been suggested in these latter works that detected variability can be explained as set by
  quake induced torsional shear oscillations of neutron star. Different aspects of this discovery  are
  currently under intense theoretical investigation (Piro 2005, Glampedakis, Samuelsson \& Andersson 2006,
  Watts \& Reddy 2007, Lee 2007; Levin 2007; Sotani, Kokkotas, \& Stergioulas 2007;
  Samuelssen \& Andersson 2007; Vavoulidis, Stavridis, Kokkotas, \& Beyer, H. 2007). An extensive survey of
  observational data and theoretical works devoted to this issue can be found in (Israel 2007; Watts and Strohmayer 2007).

  Motivated by the above interest, we focus here on the mathematical physics of the eigenfrequency problem
 for the torsional shear oscillations governed by equation of Newtonian solid mechanics. The restoring force
 is the bulk force of Hookean elastic shear stresses. Emphasis is laid on the boundary
 conditions which must be imposed on the
 toroidal field of material displacement at the edges of the seismogenic layer, that is, on the core-crust
 boundary and the star surface. Working from the homogeneous crust model we show that these boundary conditions
 substantially affect the asymptotic spectral formulae for the frequency
 of torsional shear oscillations.

  The plan of this paper is as follows. In Sec.2, a brief outline
  is given of the governing elastodynamical equation for a standard core-crust model of a quaking
  neutron star with homogeneous crust. The general
  solution of the Helmholtz equation for toroidal field of material displacements
  describing the standing-wave regime of torsional oscillations in
  the spherical polar coordinates related to the real geometry of the
  neutron  star crust is presented. Also, we obtain here the frequency spectrum for the
  nodeless global torsional elastic mode in the entire volume of the fiducial homogeneous solid star model
  which was also obtained in our
  previous investigation but by use of Rayleigh's energy method.
  In Sec.3, we derive two exact dispersion equations for the standing-wave regime of torsional
  shear vibrations corresponding to different boundary conditions at the core-crust interface and the star
  surface. A detailed analytic derivation of asymptotic spectral formulas
  is presented followed by a numerical
  analysis of the obtained frequency spectra.
  In Sec.4, we compare frequency spectra for the nodeless torsional shear oscillations
   computed from homogeneous and non-homogeneous models of the neutron star crust .
  The obtained results are summarized in Sec.5.

 \section{Governing equations}
 In two component model of quaking neutron star (Franco, Link \& Epstein 2000), its interior is thought of as
 composed of dense core (in which self-gravity is
 brought to equilibrium by the degeneracy pressure of baryon, neutron-dominated, matter) covered by highly
 conducting metal-like crustal matter composed of nuclei (basically of iron,
 $_{26}^{56}$Fe) dispersed in homogeneous Fermi-gas of relativistic electrons. The gravitational stability of the
 crust is supported by the electron degeneracy pressure $p_e$ related with the shear
 modulus as $\mu=\kappa(\rho)p_e$ (Blaes, Blandford, Madau, Koonin, 1990) with fiducial value of
 ratio\footnote{In this connection, it may be appropriate to note
 that the Local Density Approximation of microscopic theory of metals developed over the
 past three decades leads to the conclusion that transport coefficients of solid-mechanical
 elasticity, like bulk modulus and shear modulus $\mu$ are proportional to the pressure
 of degenerate Fermi-gas of conducting electrons $p_e$ squeezed between ions (Maruzzi, Janak \& Williams 1978).
 The attitude that crustal matter possesses metal-like properties lends support to the parametrization
 for shear modulus as $\mu=k(\rho)\,p_e$.} $\mu/p_e=10^{-2}$ (Strohmayer et al 1991; Cutler, Ushomirsly, Link 2003).
 It is presumed that in the approximation of continuous medium, the
 quake induced elastic deformations in crustal matter can be properly modeled by equation
 of solid mechanics for solenoidal field of material displacement $u_i$
 (e.g. Graff 1991; Lapwood \& Usami 1981; Aki \& Richards 2003)
 \begin{eqnarray}
 \label{e2.1}
 && \rho {\ddot u}_i=\nabla_k\,\sigma_{ik}\quad\quad
 \sigma_{ik}=2\mu u_{ik}\quad \\
 &&  u_{ik}=\frac{1}{2}[\nabla_i u_k+\nabla_k u_i]\quad  u_{kk}=\nabla_k  u_k=0
 \end{eqnarray}
 expressing the second law of Newtonian elastodynamics
 (McDermott, Van Horn \& Hansen 1988). The restoring force is provided
 by elastic shear stresses $\sigma_{ik}$ related to shear strains  $u_{ik}$ by Hooke
 law. Consider homogeneous model of crustal matter with constant density $\rho$ and shear modulus
 $\mu$ (e.g. Epstein 1988; Bildsten \& Cutler 1995; Franco, Link \& Epstein 2000). The shear character of
 torsional oscillations implies that they are not accompanied by fluctuations in the density: $\delta
 \rho=-\rho\,u_{kk}=0$.
 On substituting of $\sigma_{ik}$ in the equation of elastodynamics
 one has
  \begin{eqnarray}
 \label{e2.2}
 {\ddot {\bf u}}-c_t^2\,\nabla^2 {\bf u}=0\quad c_t^2=\frac{\mu}{\rho}
 \quad \nabla\cdot{\bf u}=0.
 \end{eqnarray}
 For harmonic in time fluctuations of material displacements
 \begin{eqnarray}
 \label{e2.3}
 {\bf u}({\bf r},t)={\bf a}({\bf r})\, \alpha(t)\quad\quad \alpha(t)=\alpha_0\exp(i\omega t)
 \end{eqnarray}
 substitution (\ref{e2.3}) in (\ref{e2.2}) leads to the vector Helmholtz equation
 for the standing shear wave
  \begin{eqnarray}
 \label{e2.4}
  \nabla^2{\bf u}+k^2{\bf u}=0\quad\quad \nabla\cdot {\bf u}=0\quad\quad
  k^2=\frac{\omega^2}{c_t^2}.
 \end{eqnarray}
 In what follows all calculations are carried out in the spherical coordinates with fixed polar axis
 $z$. In this frame of reference the general solution of (\ref{e2.4}) describing axisymmetric
 differentially rotational oscillations of matter in the star is
 given by the toroidal vector field (e.g. Graff 1991; Lapwood \& Usami 1981; Aki \& Richards 2003)
 \begin{eqnarray}
 \label{e2.5}
  && {\bf u}({\bf r},t)=\nabla\times {\bf r}U({\bf r},t)=\nabla U({\bf r},t)\times {\bf r}\quad U({\bf r},t)
  =f_\ell(kr)\,P_\ell(\cos\theta)\,\exp(i\omega t)\\
  \label{e2.5a}
  && f_\ell(kr)=[A_\ell\,j_\ell(kr)+B_\ell\,n_\ell(kr)]\\
  &&  \label{e2.5b} u_r=0,\,\, u_\theta=0,\,\, u_\phi=f_\ell(kr)\,P^1_{\ell}(\zeta)\exp(i\omega
  t)\\
  &&  \label{e2.5c} P^1_{\ell}(\zeta)=(1-\zeta^2)^{1/2}\frac{dP_\ell(\zeta)}{d\zeta}\,
  \quad \zeta=\cos\theta.
 \end{eqnarray}
 By $j_\ell(kr)$ and $n_\ell(kr)$ are denoted the
 spherical Bessel and Neumann functions, respectively, and by $P_\ell(\cos\theta)$ the Legendre polynomial
 of multipole degree $\ell$ (Abramowitz \& Stegun 1964).
 Function  $f_\ell$ obey the following recurrence relations
 \begin{eqnarray}
 \label{e2.6}
 &&  \frac{df_\ell}{dz}=f_{\ell-1}-\frac{\ell+1}{z}f_\ell\quad\quad
   f_{\ell-1}+f_{\ell+1}=\frac{2\ell+1}{z}f_\ell\\
   \label{e2.6a}
 && \ell
 f_{\ell-1}-(\ell+1)f_{\ell+1}=(2\ell+1)\frac{df_\ell}{dz}\quad\quad
 z=kR
 \end{eqnarray}
 which hold for both $j_\ell(kr)$ and $n_\ell(kr)$. In the
 long wavelength limit these functions are approximated by
 \begin{eqnarray}
 \label{e2.7}
 &&   j_{\ell}(z)\approx\frac{z^\ell}{(2\ell+1)!!}\quad\quad
   j_{\ell+1}(z)=\frac{z}{2\ell+3}j_\ell(z)\\
 &&  \label{e2.7a}
   n_{\ell}(z)\approx - \frac{(2\ell-1)!!}{z^{\ell+1}}\quad\quad
   n_{\ell+1}(z)=\frac{2\ell+1}{z}\,n_\ell(z).
 \end{eqnarray}
 These asymptotic formulae provide a basis for obtaining analytic estimates
 of the frequency spectra of nodeless torsion oscillations.

 As a first step, we consider global torsional oscillations in the
 entire volume of homogeneous neutron star model. For our present purpose this problem is
 interesting in that the spectral formula for the frequency of global torsional oscillations
 is used as the reference equation in testing of spectral equations for the frequency of torsional
 oscillations trapped in the crust. These latter are derived in the form
 showing that fiducial spectral formula for the global torsional oscillations
 is recovered when the core radius tends to zero.

 In the case of global torsional oscillations in entire volume
 the singular in origin solution of the Helmholtz equation for the displacement field must
 be excluded by putting $B_\ell=0$. Then, one has
 \begin{eqnarray}
 \label{e3.1}
 u_r=0,\quad u_\theta=0,\quad
 u_\phi=A_\ell\,j_\ell(kr)\,P^1_{\ell}(\zeta)\,\exp(i\omega
  t).
 \end{eqnarray}
 The standard boundary condition of stress free surface
 $n_k\,\sigma_{ik}\vert_{r=R}=0$, where $n_k$ are components of
 the unit vector normal to the star surface leads to
 \begin{eqnarray}
 \label{e3.2}
  &&n_r\sigma_{r\phi}\vert_{r=R}= \mu\left[\frac{\partial u_\phi}{\partial
  r}-\frac{u_\phi}{r}\right]_{r=R}=0\,\to\,
  \left[j_{\ell+1}(z)-\frac{j_\ell(z)}{z}(\ell-1)\right]=0 \quad z=kR.
 \end{eqnarray}
 To integrate this transcendent equation one must have recourse to
 numerical methods. In what follows we focus, however, on the long
 wavelength limit of the last equation, when $z<<1$. This is the
 regime of the nodeless torsional oscillations with the field of
 displacements $u_i$
 having no nodes in the interval $0 < r < R$; in this regime of torsional
 oscillations the components
 of $u_i$ are given by $[u_r=0,\,u_\theta=0,\,
 u_\phi={\cal A}_\ell\,r^\ell\,P^1_{\ell}(\zeta)\,\exp(i\omega t)]$.
 From this the term nodeless torsional vibrations is derived.
 The character of shear distortions in this quadrupole and octupole overtones
 of nodeless torsional oscillations is pictured pictured in Fig.1 (see, also, Bastrukov et al 2002).
 \begin{figure}
\centering{\includegraphics[width=12cm]{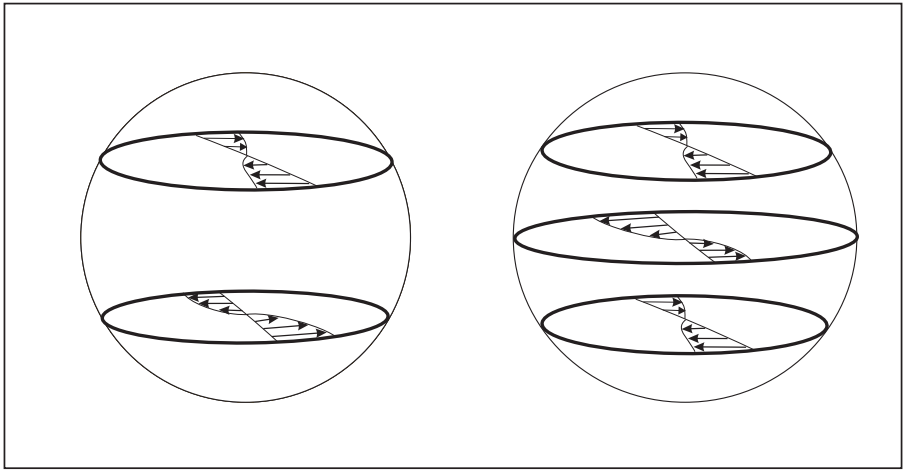}} \caption{The
 artist view of nodeless toroidal field of material displacements in
 the neutron star undergoing global torsional oscillations in
 quadrupole $(\ell=2)$ and octupole $(\ell=3)$ overtones. In
 quadrupole overtone (left), the field of material displacements in
 north and south hemispheres of neutron star undergoes out-of-phase
 oscillations, as pictured by arrows. In the octupole overtone
 (right), the material displacements in north and south are in one
 and the same phase, whereas in equatorial part the direction of
 displacements is opposite.}
\end{figure}

 In the long wavelength limit, $z<<1$, the dispersion equation (\ref{e3.2})
 is reduced, with help of approximate formulae (\ref{e2.7}) and
 (\ref{e2.7a}), to simple algebraic relation
  \begin{eqnarray}
  \label{e3.3}
   z^2-(2\ell+3)(\ell-1)=0\quad\quad z^2=k^2R^2=\frac{\omega^2}{c_t^2}R^2.
  \end{eqnarray}
 It follows that the frequency spectrum of global long wavelength
 torsional modes in a homogeneous solid star model is given by
 \begin{eqnarray}
 \label{e3.4}
 && \omega^2(_0t_\ell)=\omega_0^2[(2\ell+3)(\ell-1)]\quad\quad \omega_0^2=\frac{c_t^2}{R^2}\quad c_t^2=\frac{\mu}{\rho}\\
  \label{e3.4a}
  && \frac{\nu^2(_0t_\ell)}{\nu^2_0}=[(2\ell+3)(\ell-1)]\quad\quad
 \nu(_0t_\ell)=\frac{\omega(_0t_\ell)}{2\pi}
  \quad \nu_0=\frac{\omega_0}{2\pi}.
 \end{eqnarray}
 From now on we use standard nomenclature for the torsional
 vibration mode, as $_0t_\ell$, where zero
 marks the nodeless regime of torsional oscillations of multipole degree
 $\ell$.

 The following equivalent representations of frequency spectrum (\ref{e3.4a})
 may be useful. First, given in the form
 \begin{eqnarray}
 \label{e3.5}
  && \frac{\nu^2(_0t_\ell)}{\nu^2_0}=2(\ell+2)(\ell-1)\left[1-\frac{1}{2(\ell+2)}\right]
 \end{eqnarray}
is interesting in that at large values of multipole degree,
$\ell>>1$, can be replaced by
 \begin{eqnarray}
 \label{e3.5a}
\frac{\nu(_0t_\ell)}{\nu_0}\approx \frac{\nu'}{\nu_0}=
[2(\ell+2)(\ell-1)]^{1/2}\quad\quad \ell>>1.
 \end{eqnarray}
Second, given in the form
\begin{eqnarray}
 \label{e3.6}
  && \frac{\nu^2(_0t_\ell)}{\nu^2_0}=2\ell(\ell+1)\left[1-\frac{1}{\ell(\ell+1)}\right]
  \left[1-\frac{1}{2(\ell+2)}\right]
 \end{eqnarray}
 is represented in the limit of large $\ell$ as
 \begin{eqnarray}
 \label{e3.6a}
\frac{\nu(_0t_\ell)}{\nu_0}\approx\frac{\nu''}{\nu_0}=
[2\ell(\ell+1)]^{1/2}\quad\quad \ell>>1.
\end{eqnarray}
The comparison of ${\nu}/{\nu_0}$, ${\nu''}/{\nu_0}$ and
${\nu''}/{\nu_0}$ is shown in Fig.2; it is implied that
$\nu(_0t_\ell)=\nu$.

\begin{figure}
\centering{\includegraphics[width=12cm]{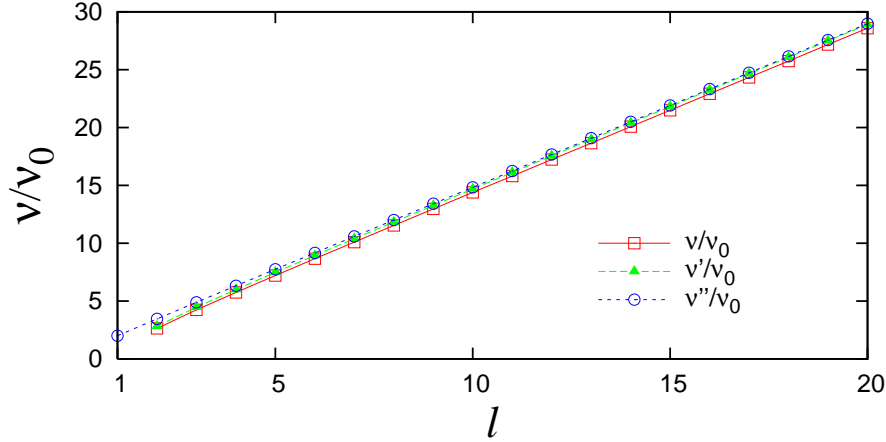}}
\caption{Fractional frequencies $\nu/\nu_0$, $\,\,\nu'/\nu_0$ and
$\nu''/\nu_0$ of global nodeless torsional oscillations as functions
of multipole degrees $1< \ell < 20$.}
\end{figure}

  It is worth noting that the spectral formula (\ref{e3.4}) can be
   derived from different
  mathematical footing, namely, by use
  of the Rayleigh's energy variational method which is particularly efficient when studying
  of non-radial nodeless oscillations of neutron stars (Bastrukov et al 1999, 2002, 2007).

  The spectral formulae like above derived are central to theoretical modal analysis
  of variability in electromagnetic emission of pulsars and
  magnetars. The works of Van Horn (1980) and Cioffi and Hansen (1980) were among the
  firsts suggested a possible association of microspikes of 10-50 milliseconds duration clearly discernable
  in the windows of main pulse train of radio pulsars with torsional oscillations of neutron stars
  (McDermott, Van Horn, Hansen 1988; Strohmayer 1991; Bastrukov et al 1999, 2007).
  This suggestion provides a guideline in the above mentioned current studies of quasiperiodic oscillations detected
  on the lightcurve tail of SGR 1806-20 and SGR 1900+14. The above
  obtained spectral formulae for the frequencies of nodeless torsional
  oscillations restored by force of shear elastic deformations have many features in common with those
  derived in (Samuelsson, Andersson 2007) from general relativistic treatment of torsional elasticity.
  The detailed identification of QPOs detected during the flare of the above magnetars with overtones
  of elastic torsional oscillations is discussed in (Watts, Strohmayer 2007).

\section{Torsional elastic vibrations in the homogeneous model of the neutron star crust}

 In the reminder of this paper we focus on torsional oscillations  trapped in the peripheral spherical
 layer of the neutron star of finite depth implying that seismically active zone depends upon the energy
 which is released in the starquake. The prime purpose of our analysis is to elucidate the effect
 of boundary conditions (reflecting the behavior of material displacements on the edges of seismogenic layer)
 on the form of frequency spectrum of elastic torsional oscillations.
 In so doing we adopt boundary conditions which are currently utilized in the
 works studying quake-induce oscillations in the neutron star crust.
 Namely the condition of stress-free-surface for both core-crust boundary and surface of the star
 and non-slip condition on the the core-crust interface (e.g. McDermott, van Horn, Hansen 1988; Strohmayer 1991;
 Bildsten \& Ushomirsky 2000).

\subsection{No-slip boundary condition on the core-crust interface
and no-stress on the star surface}
 Torsional oscillations trapped in the crust are described by
 the general solution for $u_i$ given by equations (\ref{e2.5})-(\ref{e2.6a}).
 To eliminate arbitrary constants $A_\ell$ and $B_\ell$,
 two boundary conditions, one on the core-crust interface, at
 $r=R_c$, and second on the star surface, at
 $r=R$, must be used. On the star surface we impose the standard boundary condition of the absence of
 stresses normal to surface $n_k\,\sigma_{ik}\vert_{r=R}=0$ which
 is reduced to
  \begin{eqnarray}
 \label{e4.2}
  && \mu\left[\frac{\partial u_\phi}{\partial
  r}-\frac{u_\phi}{r}\right]_{r=R}=0
 \end{eqnarray}
 and on the core-crust interface the no-slip condition
 \begin{eqnarray}
 \label{e4.3}
  && u_\phi\vert_{r=R_c}=0.
   \end{eqnarray}
 Physically, this condition means that the amplitude of differentially rotational oscillations triggered
 by starquake in the neutron star crust is gradually
 depreciated from the star surface to the core-crust interface which is the internal boundary of seismogenic
 layer. On inserting (\ref{e2.7}) in (\ref{e4.2}) and (\ref{e4.3}) we obtain
  \begin{eqnarray}
 \label{e4.4}
 &&\left[f_{\ell+1}(z)-\frac{f_\ell(z)}{z}(\ell-1)\right]=0 \quad\quad z=kR\\
 \label{e4.5}
 && f_\ell(z_c)=0\quad\quad\quad z_c=\lambda z\quad \quad  0\leq \lambda < 1.
 \end{eqnarray}
 Note, the notation $z_c=\lambda z$ means that the
 radius of the core $R_c$ can be represented as $R_c=\lambda R$ with
 $\lambda$ from the interval $0\leq \lambda < 1$. It is worth emphasizing that $\lambda$
 is strongly less than unit. In terms of Bessel
 and Neumann functions these latter  boundary conditions read
 \begin{eqnarray}
 && A_\ell\left[j_{\ell+1}(z)-\frac{j_\ell(z)}{z}(\ell-1)\right]
    + B_\ell\left[n_{\ell+1}(z)-\frac{n_\ell(z)}{z}(\ell-1)\right]=0\\
 && A_\ell\,j_{\ell}(\lambda z)+B_\ell\,n_\ell(\lambda z)=0.
 \end{eqnarray}
 Casting these equations as homogeneous matrix equation whose
 Wronskian must be equal zero we arrive at the dispersion equation
 of the form
 \begin{eqnarray}
  W(z)= \left[j_{\ell+1}(z)-\frac{j_\ell(z)}{z}(\ell-1)\right]\,n_{\ell}(\lambda z)-
  \left[n_{\ell+1}(z)-\frac{n_\ell(z)}{z}(\ell-1)\right]\,j_{\ell}(\lambda
  z)=0.
  \label{e4.8}
 \end{eqnarray}
 Computations of roots of this transcendent equation in which spherical Bessel and Neumann
 functions are defined in different points is non-trivial numerical problem
 (e.g. Pexton \& Stiger 1977). However, our prime purpose here is to evaluate the frequency spectrum
 for long wavelength differentially rotational oscillations, when $z=kR<<1$. Taking into account that
 in the long wavelength limit
\begin{eqnarray}
 \label{e4.9}
&&  j_{\ell}(z)n_{\ell}(\lambda
 z)\to \beta\,\lambda^{-(\ell+1)}\quad\quad j_\ell(\lambda
 z)n_\ell(z)\to \beta\,\lambda^\ell\quad \beta=-z^{-1}\,\frac{(2\ell-1)!!}{(2\ell+1)!!}
 \end{eqnarray}
one finds that exact dispersion equation (\ref{e4.8}) is reduced to
\begin{eqnarray}
 \label{e4.10}
  && z^2=(2\ell+3)[(\ell-1)+(\ell+2)\lambda^{2\ell+1}]\quad\quad
  z^2=k^2R^2=\frac{\omega^2}{c_t^2}R^2.
 \end{eqnarray}
From this it follows
\begin{eqnarray}
 \label{e4.10a}
  && \frac{\nu^2_1}{\nu^2_0}=
  (2\ell+3)(\ell-1)\left[1+\frac{\ell+2}{\ell-1}\lambda^{2\ell+1}\right]\\
  && \nu_1=\frac{\omega_1}{2\pi}\quad
  \nu_0=\frac{\omega_0}{2\pi}\quad
  \lambda=\frac{R_c}{R}=1-h\quad h=\frac{\Delta R}{R}.
 \label{e4.11}
 \end{eqnarray}
 In the limit of zero-size radius of the core, $\lambda=(R_c/R)\to
 0$, corresponding to torsional oscillations in the entire volume of a solid star,
 we regain spectral equation (\ref{e3.4a}) for the frequency of global torsional mode.
 For our further purpose we note that equation (\ref{e4.10a}) can be
 represented in the following equivalent form
 \begin{eqnarray}
 \label{e4.12}
  && \frac{\nu_1}{\nu_0}=
  [(\ell+2)(\ell-1)]^{1/2}p_1^{-1}\quad\quad
 p_1^{-1}=\left[2\left(1-\frac{1}{2(\ell+2)}\right)
 \,\left(1+\frac{\ell+2}{\ell-1}\lambda^{2\ell+1}\right)\right]^{1/2}
 \end{eqnarray}
 which is discussed in the next subsection.
 Henceforth the suffice in expressions for $\nu$ and $\omega$ marks the number of eigenfrequency
 problem under consideration. The considered in this subsection is marked by suffice 1,
 and two another conceivable boundary conditions are regarded in the reminder of this work.

\subsection{Stress free boundary conditions on both core-crust interface and
the neutron star surface}
 Now we adopt the boundary conditions of the free from stresses surfaces
 for both the surface of the star and the core-crust boundary
 $n_k\,\sigma_{ik}\vert_{r=R,R_c}=0$:
  \begin{eqnarray}
 \label{e5.1}
  && n_r\sigma_{r\phi}\vert_{r=R,R_c}=
  \mu\left[\frac{\partial u_\phi}{\partial
  r}-\frac{u_\phi}{r}\right]_{r=R,R_c}=0
 \end{eqnarray}
 which in terms of $f_\ell(z)$ are given by
  \begin{eqnarray}
 \label{e5.2}
   &&
   \left[\frac{df_\ell(z)}{dz}-\frac{f_\ell(z)}{z}\right]_{z=kR}=0\quad\quad
   \left[\frac{df_\ell(z_c)}{d z_c}-\frac{f_\ell(z_c)}{z_c}\right]_{z_c=\lambda kR}=0\quad\quad  0\leq \lambda <1.
 \end{eqnarray}
For a non-vanishing solution of these equations to exist, the
following dispersion equation
\begin{eqnarray}
\nonumber
 &&\left[j_{\ell+1}(z)-\frac{j_\ell(z)}{z}(\ell-1)\right]\left[n_{\ell+1}(\lambda z)-
  \frac{n_\ell(\lambda z)}{\lambda z}(\ell-1)\right]\\
 &&  -
 \left[j_{\ell+1}(\lambda z)-\frac{j_\ell(\lambda z)}{\lambda
 z}(\ell-1)\right] \left[n_{\ell+1}(z)-\frac{n_\ell(z)}{z}(\ell-1)\right]=0
 \label{e5.3}
 \end{eqnarray}
must hold. Following the line of argument of foregoing subsection,
consider the limit of long wavelengths, $z=kR<<1$. This yields
 \begin{eqnarray}
  \label{e5.5}
 && \frac{\nu^2_2}{\nu^2_0}= (2\ell+3)(\ell-1)
 \,\left[\frac{1+\lambda^{2\ell+1}}{1-\lambda^{2\ell+3}}\right]\quad \quad 0\leq \lambda <1.
 \end{eqnarray}
 In the limit $\lambda=(R_c/R)\to
 0$, the last equation is reduced to spectral formula (\ref{e3.4a}) for the frequency of global torsional
 oscillations.

\begin{figure}
\centering{\includegraphics[width=12cm]{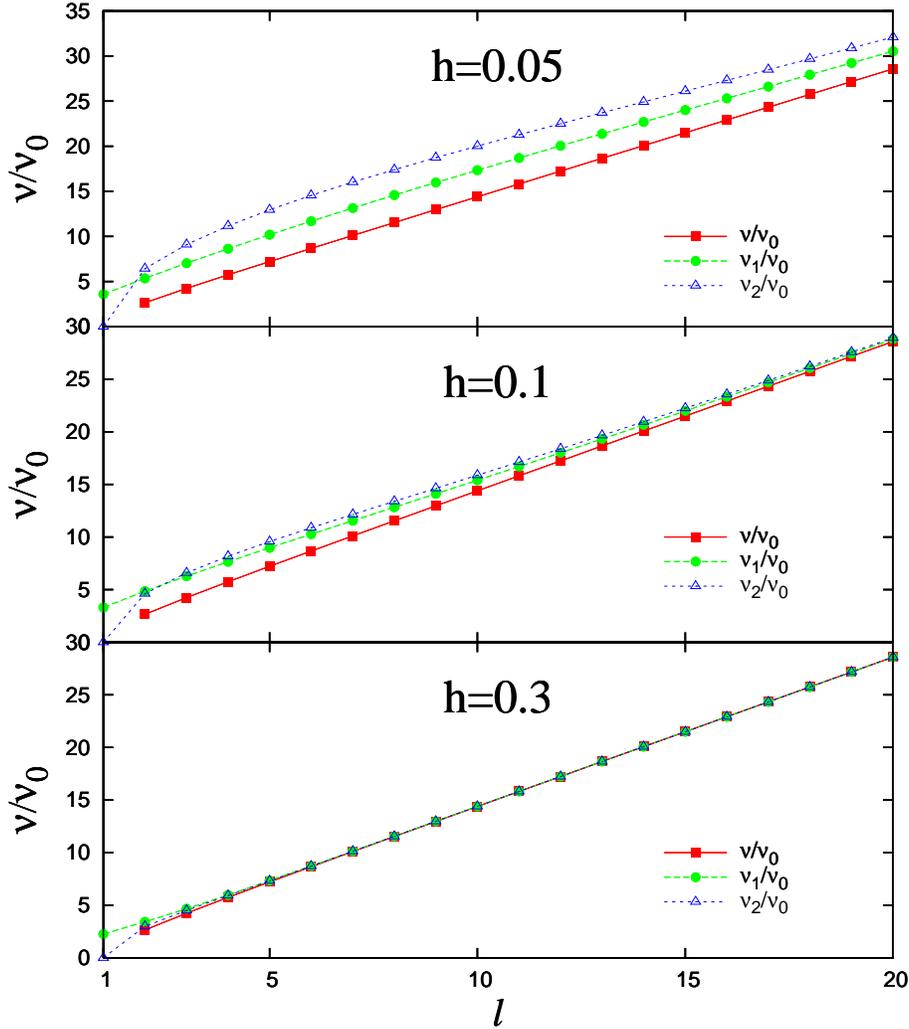}}
\caption{Fractional frequencies of nodeless torsional shear
oscillations as functions of multipole degrees $1< \ell < 20$
plotted at $h=\Delta R/R=0.05, 0.1$ and $0.3$. $\nu/\nu_0$ for
global nodeless torsional mode in entire volume of homogeneous
neutron star model; $\nu_1/\nu_0$ and $\nu_2/\nu_0$ for the
frequency of torsional modes trapped in the peripheral layer of
homogenous model of crust computed with boundary conditions of first
and second examples.}
\end{figure}

 Fig.3 illustrates the general trends of fractional frequencies - $\nu/\nu_0$,  $\nu_1/\nu_0$ and  $\nu_2/\nu_0$
 as functions of multipole degrees $1< \ell < 20$ plotted at the values of fractional depth given by
 $h=\Delta R/R=0.05, 0.1$ and $0.3$.
 The difference caused by boundary conditions is notable in the low-$\ell$ domain, when the depth of seismogenic
 layer  $\Delta R=0.5$ and $1.0$ km (in the neutron star of fiducial radius $R=10$ km).
 This difference is practically vanishes when the thickness of layer is
 about $\Delta R=3.0$ km
 and larger. Generally, the thicker this depth the less difference.
 Also we note, for our further purpose, that equation (\ref{e5.5}) can be represented as
\begin{eqnarray}
 \label{e5.6}
  && \frac{\nu_2}{\nu_0}= [(\ell+2)(\ell-1)]^{1/2}p_2^{-1}\quad\quad p_2^{-1}=\left[2\left(1-\frac{1}{2(\ell+2}\right)
 \,\frac{1+\lambda^{2\ell+1}}{1-\lambda^{2\ell+3}}\right]^{1/2}.
 \end{eqnarray}

 At this point it seems worthy of noting that analogous, from mathematical side, problem of torsional
 vibration mode trapped in the peripheral solid layer has been considered long ago by Pekeris (1965) in the geoseismic
 context. The  obtained in this latter work spectral formula
 with boundary conditions identical to considered in this subsection reads
 \begin{eqnarray}
 \label{e5.7}
 && \frac{\nu_p}{\nu_0}=
 [(\ell-1)(\ell+2)]^{1/2}p^{-1}_{\ell}\quad\quad \ell\geq 10.
 \end{eqnarray}
 In work of Pekeris (1965) the dimensionless parameter $p_\ell$, has been numerically computed from
 integral equation establishing "relation between geometrical optics and terrestrial spectroscopy" and
 tabulated for $1\leq \ell\leq 50$.  Remarkably, in appearance the Pekeris spectral formula is identical to the
 above presented equations (\ref{e4.12}) and (\ref{e5.6}).
 In Fig.4 we plot $\nu_p/\nu_0$ with use of data for $p_\ell$
 taken from (Pekeris 1965, Table 2) as a function of $\ell$ in
 juxtaposition with our spectral formula (\ref{e5.7}) in which we have used the value of fractional depth
 of seismogenic layer given by $h=0.1$. Analogous conclusion holds for $\nu_1/\nu_0$.
 It should be noted, however, that equation (\ref{e5.7}) has been obtained
 with use of non-uniform profile for the speed of transverse wave of elastic
 shear $c_t(r)$ which is used as input function of the method, not computed.
 In view of this, the last figure can be considered no more than juxtaposition, not a
 comparison, of our and Pekeris spectral equations. Nonetheless, the fact that
 both approaches yields practically
 identical results suggests that presented in this section analysis can be utilized for assessing frequencies of
 torsional seismic vibration modes in the solid Earth-like planets too.

 \begin{figure}
 \centering{\includegraphics[width=10cm]{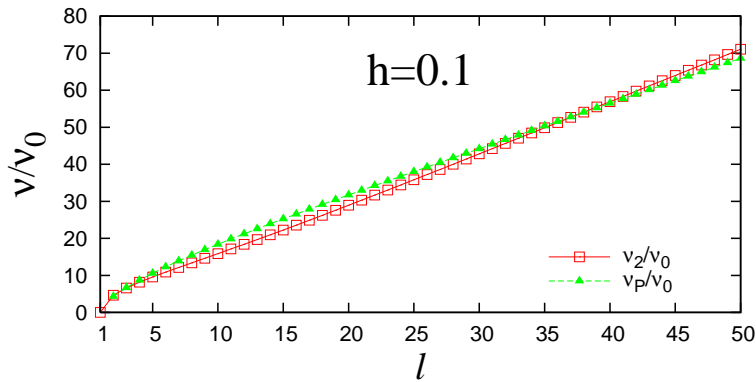}}
\caption{Frequency of torsional nodeless vibrations computed in the
homogeneous model of peripheral seismic layer of second example of
this work in juxtaposition with Pekeris (1965) asymptotic spectral
formula for torsional geoseismic vibrations.}
\end{figure}

\section{Nodeless torsional oscillations in homogeneous and non-homogeneous models of the neutron star crust}
 From above it follows that in the long wavelength limit
 the toroidal field of displacements (\ref{e2.5})-(\ref{e2.6a}) is reduced to the form that can be conveniently
 represented as follows
 \begin{eqnarray}
 \label{e6.1}
  && {\bf u}({\bf r},t)={\bf a}({\bf r})\,\alpha(t) \quad \alpha(t)=\alpha_0\exp(i\omega t)\\
  \label{e6.2}
  && {\bf a}({\bf r})=\nabla \chi({\bf r})\times {\bf r}\quad
  \chi({\bf r})=f_\ell({\bf r})P_\ell(\zeta)\\
   \label{e6.2a}
  && f_\ell({\bf r})=[{\cal A}_\ell\,r^\ell+{\cal
  B}_\ell\,r^{-(\ell+1)}].
 \end{eqnarray}
 The toroidal field (\ref{e6.1})-(\ref{e6.2a}) is the general solution to the vector Laplace equation
 \begin{eqnarray}
  \label{e6.3}
  \nabla^2\,{\bf u}({\bf r},t)=0\quad\quad \nabla{\bf u}({\bf
  r},t)=0.
 \end{eqnarray}
 which can be thought of as the long wavelength limit of the Helmholtz
 equation $\nabla^2{\bf u}+k^2{\bf u}=0$, because in the limit of long
 wavelengths, $\lambda\to \infty$, the wave vector $k=(2\pi/\lambda)\to 0$.
 The vector Laplace equation can be regarded
 as fundamental equation defining regime
 of nodeless shear vibrations \footnote{The axisymmetric, odd parity, toroidal vector field
 represents one of two fundamental (mutually
 orthogonal and different in parity) solutions to (\ref{e6.1}) describing nodeless torsional
 vibrations. The second fundamental solution is given by  even parity poloidal vector field
 (Bastrukov et al 2007) which describe nodeless spheroidal vibration
  mode, in accord with canonical Lamb's classification of vibrational modes in an elastically deformable solid sphere
 (e.g. Lapwood \& Usami 1981; Aki \& Richards 2003).}.
  The practical usefulness of this attitude is that it allows to assess the difference
  between predictions for the frequency spectra computed on equal footing within homogeneous and inhomogeneous models
  of the neutron star crust, that is, regardless of the form of density and shear modulus profiles.
  Most efficiently it can be done by Rayleigh's energy method (Bastrukov et al 2007).
 The point of departure is the integral equation of
 energy conservation which is obtained from equation of elastodynamics (\ref{e2.1}) by
 its scalar multiplication with $u_i$ and integration over the
 crus volume
 \begin{eqnarray}
 \label{e6.8a}
  \frac{\partial }{\partial t}\int \frac{\rho {\dot u}^2}{2}\,d{\cal
  V} = -\int \sigma _{ik}{\dot u}_{ik}\,d{\cal V}= -2\int \mu\, u_{ik}{\dot u}_{ik}d{\cal
  V}.
   \end{eqnarray}
 On inserting in (\ref{e6.8a}), the separable form of the displacement field (\ref{e6.1}),
 we arrive at equation for temporal amplitude $\alpha(t)$ having the form of
 the well-familiar equation of normal vibrations
 \begin{eqnarray}
 \label{e6.8b}
 && \frac{dE}{dt}=0\quad E=\frac{M{\dot\alpha}^2}{2}+\frac{K{\alpha}^2}{2}
 \\
 && {\ddot\alpha}+\omega^2\alpha=0\quad\quad \omega^2=\frac{K}{M}\\
 && M=\int \rho(r)\, a_i\,a_i\,d{\cal V}\quad\quad
 K=2\int \mu(r)\, a_{ik}\,a_{ik}\,d{\cal V}\quad  \quad a_{ik}=\frac{1}{2}[\nabla_i a_k + \nabla_k
 a_i].
 \end{eqnarray}
 The analytic form of the inertia $M$ and stiffness
 $K$ shows that method can be utilized for computing frequency $\omega^2=K/M$  of shear vibrations
 (both spheroidal and torsional) for wide class of models with non-uniform density and shear modulus
 profiles in the crust which are the input parameters of the method.
 Taking the integral over the solid angle in the above expression for inertia $M$
 we obtain
 \begin{eqnarray}
 \label{e6.8c}
 M=\frac{4\pi\,\ell(\ell+1)}{2\ell+1} {\cal A}_\ell^2 \left[
 \int\limits_{R_c}^{R}\rho(r)\,r^{2\ell+2} dr+\frac{2{\cal B}_\ell}{{\cal A}_\ell} \int\limits_{R_c}^{R} \rho(r)\,r
 dr+ \frac{{\cal B}_\ell^2}{{\cal A}_\ell^2} \int\limits_{R_c}^{R}
 \rho(r)\,r^{-2\ell} dr\right].
\end{eqnarray}
In similar fashion, for the parameter of rigidity $K$ we get
 \begin{eqnarray}
 \label{e6.8d}
 K= 4\pi\, {\cal A}_\ell^2(\ell^2-1)\left[
 \int\limits_{R_c}^{R}\,\mu(r)\,r^{2\ell}dr+
  \frac{{\cal B}_\ell^2}{{\cal A}_\ell^2} \frac{\ell(\ell+2)}{(\ell-1)}\int\limits_{R_c}^{R}\mu(r)\,r^{-2\ell-2}dr\right].
\end{eqnarray}
 These later equations for $M$ and $K$ emphasizes the fact that the choice of
 boundary conditions does matter for the problem under consideration.

 As a representative example, consider torsional oscillations in
 the crust with constants ${\cal A}_\ell$ and ${\cal B}_\ell$ eliminated from
 the following boundary conditions
  \begin{eqnarray}
  \label{e6.5}
 && u_\phi\vert_{r=R_c}=0\quad u_{\phi}\vert_{r=R}=[\mbox{\boldmath $\phi$}_R\times {\bf R}]_\phi \\
 && \mbox{\boldmath $\phi$}_R=\exp(i\omega t)\nabla_{\hat{\bf n}} P_\ell(\zeta)\quad\quad
 \nabla_{\hat{\bf n}}=\left(0,\frac{\partial }{\partial
 \theta},\frac{1}{\sin\theta}\frac{\partial }
 {\partial \phi}\right).
 \end{eqnarray}
 First is the no-slip condition on the core-crust interface, $r=R_c$, and
 the form of second boundary condition on the star surface, at $r=R$,
 is dictated by symmetry of general toroidal field of nodeless torsional oscillations.
 The resultant algebraic equations steaming from boundary conditions (\ref{e6.5})
 lead to
  \begin{eqnarray}
   \label{e6.6}
 {\cal A}_\ell={\cal N}_\ell\quad {\cal B}_{\ell}=-{\cal
 N}_\ell\,R_c^{2\ell+1}\quad\quad {\cal
 N}_\ell=\frac{R^{\ell+2}}{R^{2\ell+1}-R_c^{2\ell+1}}.
 \end{eqnarray}
 In the homogeneous crust model, presuming constant values
 of $\rho$ and $\mu$, we get
 \begin{eqnarray}
 \nonumber
 && M(\ell,\lambda)=\frac{4\pi\ell(\ell+1)}{(2\ell+1)(2\ell+3)}\frac{\rho
 R^5}{(1-\lambda^{2\ell+1})^2}\\
 &&\times
 \left[1-
 (2\ell+3)\lambda^{2\ell+1}+\frac{(2\ell+1)^2}{2\ell-1}\lambda^{2\ell+3}-
 \frac{2\ell+3}{2\ell-1}\lambda^{2(2\ell+1)}\right]\\
 && K(\ell,\lambda)=\frac{4\pi\ell(\ell^2-1)}{2\ell+1}\,\frac{\mu R^{3}}{(1-\lambda^{2\ell+1})}
 \left[1-\frac{(\ell+2)}{(\ell-1)}\lambda^{2\ell+1}\right].
\end{eqnarray}
 Note, in the limit $\lambda\to 0$ we again recover the spectral
 formula for the frequency $\omega^2=K/M$ of the global torsional oscillations in the entire
 volume of the homogeneous solid star model (\ref{e3.4a}).

\subsection{Numerical analysis}

 In the above expanded the energy variational method of computing the frequency of elastic
 shear oscillations, the profiles of density $\rho(r)$ and pressure
 $p(r)$ linearly proportional to the shear modulus $\mu(r)$ of crustal matter
 are regarded the input parameters. The approach we adopt is the one used by previous authors taking these parameters
 from evolution models aimed at computing gravitationally equilibrium state of matter and the internal structure
 of neutron star. In these later models the density and pressure are computed
 from Tolman-Oppenheimer-Volkoff (TOV) equation with account for realistic EOS
 (e.g. Weber 1999; Lattimer \& Prakash 2001).

 \begin{figure}
\centering{\includegraphics[width=8cm]{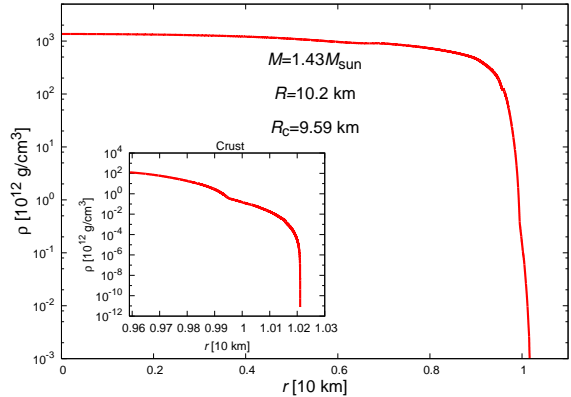}}
\caption{Density profile for realistic non-homogeneous neutron star
model built on the TOV equation and EOS given in (Wiringa et al
1988). Insertion depicts density distribution in the peripheral
crustal region.}
\end{figure}

\begin{figure}
\centering{\includegraphics[width=8cm]{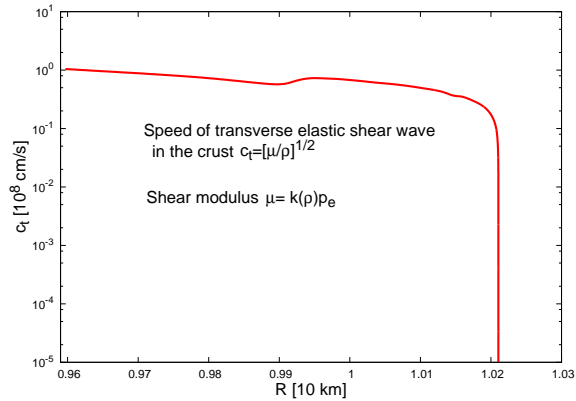}}
 \caption{Profile of speed of transverse shear wave in
the crust of non-homogeneous model with density pictured in
Fig.4.}
 \end{figure}

 Specifically, we adopt here the model of neutron star model with
 fiducial mass of $M=1.43M_{sun}$ presented in Wiringa, Fiks \& Fabrocini (1988)
 and will use parameters for the crust matter given in Douchin \& Haensel (2001).
 The density profiles in the star as a whole and in the crust of this star model
 are pictured in Fig.5.
 The position of the core-crust interface is placed at the fiducial density $\rho=1.5\, 10^{14}$ g cm$^{-3}$,
 in accord with arguments of works (Cutler, Ushomirsky, Link 2003; Pethick, Ranenhall \& Lorentz 1995).
 The effect of non-homogeneous density in the crust on the speed of transverse wave of elastic shear in this model is
 illustrated in Fig.6.

 In Fig.7 we compare the fractional frequency of torsional
 oscillations trapped in the crust of fixed depth $\Delta R=0.6$ km (that has been normalized to the value of
 $\nu_0=15$ Hz) computed in the homogeneous and inhomogeneous crust
 models. It is seen that
 prediction of homogenous crust model with boundary conditions of this section are quite different from
 that inferred in previous section from homogeneous model too, but for different boundary conditions.
 Also, the predictions of homogeneous models
 are drastically different from those for non-homogenous one.
 This difference is manifested in both the overall trends of frequency as a function of multipole
 degree and absolute values of fractional frequencies. One of sources of uncertainties
 is parametrization of shear modulus profile. In the curve computed with $\mu=k(\rho)p_e$
 we used parametrization of work (Cutler, Ushomirsky, Link 2003) and with $\mu=c_t^2\rho$ from
 (Strohmayer et al 1991). At this point we leave the discussion of mathematical details of numerical
 calculations because this issue is not the main subject of presented investigation.

 \begin{figure}
 \centering{\includegraphics[width=10cm]{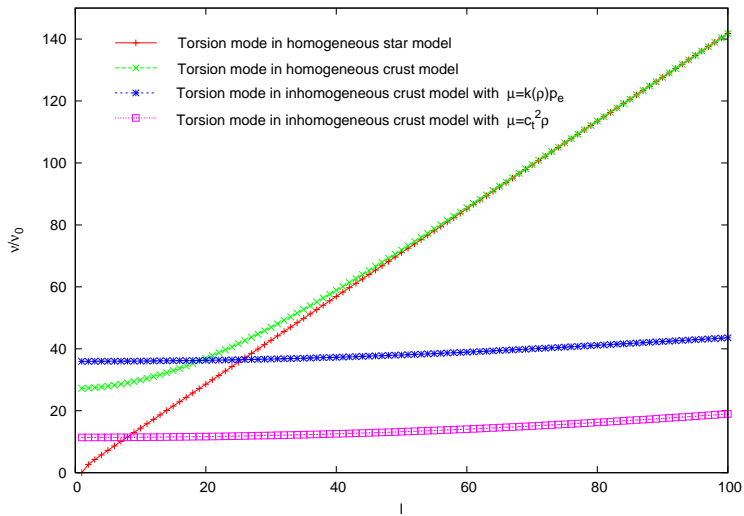}}
 \caption{Frequency of nodeless torsional shear oscillations as a
 function of multipole degree computed in the homogeneous and
 inhomogeneous models.}
 \end{figure}

\section{Summary}
 There are now quite solid arguments showing that the Soft Gamma-Ray Repeaters are isolated,
 non-accreting, seismically active magnetars  (Thompson \& Duncan 1995) -- quaking neutron stars endowed
 with ultrastrong magnetic fields. The X-ray bursting luminosity of magnetar is associated
 with starquakes which are thought of as sudden release of magnetic field stresses cracking of
 the neutron star crust by the X-ray flares. Evidence in favor of seismic nature of the magnetar
 flares is provided by striking similarities between statistics of SGR's bursts and earthquakes
 (Cheng,  Epstein, Guyer \& Young 1996). The common belief is that the dynamics of quake induced internal elastic and
 magnetic field stresses in the crust can be properly understood
 within the framework of the two-component core-crust model of quaking neutron stars (Franco, Link \& Epstein
 2000), provided that the core-crust coupling is dominated by an ultrastrong magnetic field.
 A conceivable and comprehensive, from a physical point of view, explanation of the magnetic
 core-crust coupling provides a model of paramagnetic neutron star (Bastrukov et al 2002; Bastrukov et al  2003).
 In  this model the neutron star core is regarded as a spherical
 bar magnet (composed of baryon matter dominated by
 poorly conducting neutron component) permanently magnetized to saturation
 due to Pauli's mechanism of field-induced paramagnetic spin-polarization of neutron magnetic moments.
 The microstructure and extremely high conductivity of crustal matter indicates to
 its metal-like electrodynamical properties. This analogy suggests that
 the magnetic coupling of the core (permanent magnet) with the crust (metal) is
 similar to the well-known magnetic adhesion between a metal specimen and bar magnet.
 The fast process of postquake recovery of the magnetar is of course dominated by
 force of gravitational pull. The prime effect of magnetic field frozen in the core on this process
 is that the the perturbed by starquake a highly conducting crustal matter
 (as well as the less dense plasma of magnetar corona (Beloborodov \& Thompson 2007)
 expelled from the star surface by SGR's flares) sets in axisymmetric torsional
 oscillations about axis of ultrastrong magnetic field frozen in the star.
 It is these oscillations about magnetic axis are observed, as is argued in (Israel 2007; Watts, Strohmayer 2007)
 as quasiperiodic oscillations of x-ray luminosity with high signal-to-noise
 ratio. In (Watts \& Strohmayer 2007) arguments are given
 that the data favor the idea of a seismic origin of the detected QPOs, that is, as caused by quake
 induced torsional shear oscillations of crustal matter of magnetar.
 Adhering to this attitude, we have investigated
 several scenario of torsional shear vibrations restored by bulk
 force of shear elasticity (the problem from which the very notion of
 torsion shear vibrations of an elastic sphere came into existence).
 Our prime purpose  was to elucidate the distinctions between spectral
 formulae for the frequency of nodeless torsional shear oscillations caused by
 different behavior of quake induced crustal matter on the boundaries of
 the crust.  In pursuing this goal and using canonical methods of mathematical physics
 we have considered several examples of exact solutions of the eigenfrequency
 problem demonstrating crucial effect of boundary conditions on the frequency
 spectrum of torsional mode of elastic oscillations trapped in the crust.
 The asymptotic spectral formulae for the frequency of nodeless torsion oscillations
 have been presented so that they can be conveniently applied to a wide
 class of celestial objects.
 The practical usefulness of the exact solutions
 for the toroidal field of material displacements considered here is
 that they can be utilized in the study of torsional shear vibrations restored by forces of intrinsic stresses
 of different physical nature, like Newtonian gravitation field stresses (e.g. Shu 1992) and
 Maxwellian magnetic field stresses (e.g. Chandrasekhar 1961; Mestel 1999), not
 only Hookean elastic one. With all that we conclude, while the obtained spectral formulae properly match the
 observable QPOs frequency, the information obtained from only these models is less than necessary to draw definite
 physical
 statements which of above forces plays dominant role. The point of particular interest, as is generally believed,
 are torsion oscillations driven by the Lorentz force that can be represented as divergence of fluctuating
 magnetic field stresses (Franco, Link, Epstein 2000)
 \begin{eqnarray}
 \nonumber
 && \delta T_{ik}=\frac{1}{4\pi}[B_i\,\delta B_k+B_k\,\delta
 B_i-B_j\,\delta
 B_j\delta_{ik}]\\ \nonumber
 &&  \delta {\bf B}=\nabla\times [{\bf u}\times {\bf B}]\quad\quad \nabla\cdot {\bf u}=0.
  \end{eqnarray}
 The canonical form of magneto-solid-mechanical
 equations governing Alfv\'enic, compression free, oscillations in a perfectly conducting elastically deformable solid,
 regarded as material continuum, reads
 \begin{eqnarray}
 \nonumber
 \rho {\ddot u}_i=\nabla_k\,\delta T_{ik}\quad \quad\quad
  \frac{\partial }{\partial t}\int \frac{\rho {\dot u}^2}{2}\,d{\cal
  V} = -\int \delta T_{ik}{\dot u}_{ik}\,d{\cal V}.
 \end{eqnarray}
 The integrand of equation of energy conservation exhibits
 the fact that the work done by magnetic field stresses in the volume of a quaking neutron star is accompanied
 too by shear deformations and, thus, shows the torsional shear oscillations can also be sustained by
 fluctuations of the magnetic field stresses.
 We shall turn to this problem in a forthcoming paper.

 The authors are indebted to Dr. Judith Bunder (UNSW, Sydney) for critical reading of the manuscript and
 valuable comments. This work is partly supported by NSC of Taiwan, Republic of China, under grants
 NSC 96-2811-M-007-012 and NSC 96-2628-M-007-012-MY3.

\end{document}